\documentstyle[aps,amssymb]{revtex}

\newcommand{\qs}{\left[}

\newcommand{\qd}{\right]}

\newcommand{\nn}{\nonumber}
\newcommand{\x}{{\bf x }}

\newcommand{\de}{{{\mathrm{d}}}}

\newcommand{\kM}{$\kappa$-Minkowski}

\newcommand{\R}{{\mathbb{R}}}

\def\be{\begin{equation}}
\def\ee{\end{equation}}
\def\bea{\begin{eqnarray}}
\def\eea{\end{eqnarray}}

\makeatletter
\@addtoreset{equation}{section} %%% Latex Companion, p.~16
\makeatother

\def\appendix#1
{
\addtocounter{section}{1}\setcounter{equation}{0}%\setcounter{section}{1}
 \renewcommand{\thesection}{\Alph{section}}
 \section*{%Appendix~
 \thesection\protect\indent \parbox[t]{16.715cm}{#1}}
 \addcontentsline{toc}{section}{Appendix \thesection\ \ \ #1}
}

\def\<#1,#2>{\left\langle#1,#2\right\rangle} %% bilinear form
\def\nn{\nonumber}

\def\be{\begin{equation}}
\def\ee{\end{equation}}
\def\bea{\begin{eqnarray}}
\def\eea{\end{eqnarray}}

\newcommand{\kP}{$\kappa$-Poincar\'{e}}
\newcommand{\nab}[1]{\vec{x}\,\vec{\nabla}}

\begin{document}

% \title{\begin{flushright}
% hep-th/0306013 \\
% $~$
% %{\it astro-ph/0407227 \\
% %$~$ \\
% %November 2002}
% \end{flushright}

\title{{\Large {\bf Action functional for \kM\ Noncommutative Spacetime}}}

\author{$~$\\
{\bf Alessandra~AGOSTINI}$^a$, {\bf Giovanni~AMELINO-CAMELIA}$^b$,
{\bf Michele ARZANO}$^c$
and {\bf Francesco~D'ANDREA}$^d$}
\address{$^a$Dipartimento di Scienze Fisiche,
Universit\`{a} di Napoli ``Federico II'',\\
    Monte S.~Angelo, Via Cintia, 80126 Napoli, Italy\\
$^b$Dipartimento di Fisica, Universit\`{a} di Roma ``La Sapienza''
and INFN Sez.~Roma1,\\
    P.le Moro 2, 00185 Roma, Italy\\
$^c$Institute of Field Physics, Department of Physics and Astronomy,\\
    University of North Carolina, Chapel Hill, NC 27599-3255, USA\\
$^d$Mathematical Physics Sector, International School for Advanced
Studies (SISSA), \\
    Via Beirut 2-4, Trieste, Italy}
%$~~~~~~~~$ $~~~~~~~~$ $~~~~~~~~$ $~~~~~~~~$ $~~~~~~~~$}

\maketitle

\begin{abstract}\noindent%
We examine some
alternative possibilities for an action functional for $\kappa$-Minkowski
noncommutative spacetime, with an approach which should be applicable
to other spacetimes with coordinate-dependent commutators of
the spacetime coordinates ($[x_\mu,x_\nu]=f_{\mu,\nu}(x)$).
Early works on $\kappa$-Minkowski focused
on $\kappa$-Poincar\'e covariance
and the dependence of the action functional on
the choice of Weyl map, renouncing to invariance under cyclic permutations of
the factors composing the argument of the action functional.
A recent paper (hep-th/0307149), by
Dimitrijevic, Jonke, Moller, Tsouchnika, Wess and Wohlgenannt,
focused on a specific choice of Weyl map and,
setting aside the issue of $\kappa$-Poincar\'e
covariance of the action functional, introduced in implicit form
a cyclicity-inducing measure. We
provide an explicit formula for (and derivation of) a choice of measure which
indeed ensures cyclicity of the action functional,
and we show that the same choice of measure is
applicable to all the most used choices of Weyl map. We find that
this ``cyclicity-inducing measure'' is not covariant
under $\kappa$-Poincar\'e transformations.
We also notice that the
cyclicity-inducing measure can be straightforwardly derived
using a map which connects the $\kappa$-Minkowski spacetime
coordinates and the spacetime
coordinates of a ``canonical'' noncommutative spacetime,
with coordinate-independent commutators.
\end{abstract}

\bigskip
\bigskip

\date{\today}
%\preprint{hep--th/0302NNN}

\maketitle

\newpage
%%% ======================================================================
\section{Introduction}\noindent
A sizeable literature has been devoted to noncommutative versions of Minkowski
spacetime. Certain types of arguments~\cite{dopl1994}
for an ``uncertainty principle
for localization'' lead one to consider the simplest such spacetimes, the
``canonical noncommutative spacetimes''
\begin{equation}\label{canonical}
[{\x}_\mu,{\x}_\nu]= i\theta_{\mu,\nu}
~.
\end{equation}
Other possible forms of the
uncertainty principle for localization may lead~\cite{kpoinplb} to a
``Lie-algebra'' form of noncommutativity:
\begin{equation}\label{liealg}
[{\x}_\mu,{\x}_\nu]= i \zeta_{\mu,\nu}^{\sigma} {\x}_\sigma
~.
\end{equation}
(Both $\theta_{\mu,\nu}$ and $\zeta_{\mu,\nu}^{\sigma}$
are coordinate-independent.)

It is emerging that the canonical spacetimes may play a
role~\cite{dougnekr} in an effective-theory description of
the physics of strings in presence of a spacetime-coordinate-indepedent
$B$-tensor external background. For appropriate choice of spacetime dependence
of the $B$-tensor external background one can instead
obtain~\cite{cerchi,orfeuLieAlg} a description in terms of a Lie-algebra
spacetime. Recent results~\cite{kodadsr,jurekkodadsr} suggest that Lie-algebra
spacetimes may also play a role in the description of some formulations of
Loop Quantum Gravity.

These results have motivated a strong interest in the construction of field
theories in noncommutative spacetimes. For the case of the simple canonical
spacetimes there are already some field-theory proposals for which a rather
advanced level of development has been achieved (see, {\it e.g.},
Refs.~\cite{dougnekr}). But for the case of Lie-algebra
spacetimes (and in general of spacetimes with coordinate-dependent commutators
of the spacetime coordinates $[x_\mu,x_\nu]=f_{\mu,\nu}(x)$) several additional
difficulties are encountered~\cite{lukieFT,gacMajidgacMichele,aad03} in the
attempts to construct field theories. The difficulties start already at the
level of constructing an action functional, which should be a map from
functions of the noncommutative spacetime
coordinates to the complex numbers. It appears that one
should renounce to some familiar properties that the action functional enjoys
in commutative spacetimes, and it is difficult to choose which properties
should be maintained and which one should be given up.

These issues concerning the action functional for Lie-algebra noncommutative
spacetimes have been most extensively considered for the case of the ``\kM ''
noncommutative spacetime''~\cite{MajidRuegg,lukieAnnPhys}, whose coordinates
satisfy the commutation relations
\begin{equation}\label{eq:kM}
[{\x}_j,{\x}_0]=i\lambda {\x}_j~,~~~[{\x}_j,{\x}_k]=0
\end{equation}
where $j,k=1,2,3$ and we denote the dimensionful
noncommutativity parameter by $\lambda$
(other \kM\ studies introduce the parameter $\kappa$, which is related to the
$\lambda$ of (\ref{eq:kM}) by $\lambda=\kappa^{-1}$). In most early works on
$\kappa$-Minkowski the action functional was structured in such a way to
reflect fully the underlying $\kappa$-Poincar\'e
invariance~\cite{lukieFT,gacMajidgacMichele,aad03,jurekKMKP}, even allowing for
a possible dependence on the choice of Weyl map (a choice which affects the
description of the $\kappa$-Poincar\'e transformations). However, this
invariance criterion was found to lead to action functionals
which do not enjoy
invariance under cyclic permutations of the fields
\begin{equation}\label{noncycl}
\int\, f_1(\x) f_2(\x) \dots f_{n-1}(\x)f_n(\x) \neq \int\, f_n(\x) f_1(\x)
f_2(\x) \dots f_{n-1}(\x) ~.
\end{equation}
Without cyclicity of the action functional many familiar features of field
theory are immediately lost and some apparently unsurmountable difficulties are
encountered. In the recent Ref.~\cite{wess0307149} Dimitrijevic, Jonke, Moller,
Tsouchnika, Wess and Wohlgenannt, while focusing
on a specific choice of Weyl map,
considered a possible cyclic action functional. The corresponding integration
measure was not described explicitly in Ref.~\cite{wess0307149}, and its
possible dependence on the choice of Weyl map was not considered. Moreover, the
issue of $\kappa$-Poincar\'e invariance of the action functional, which had
been central in previous works, was not explored in Ref.~\cite{wess0307149}.

Here we provide explicitly a choice of measure which indeed ensures cyclicity
of the action functional, and we find that the same choice of measure is
applicable to all the most common choices of Weyl map. We show that this choice
of measure can be derived constructively using only $\kappa$-Minkowski
properties, but it can also be derived more simply by exploiting a map which
exists between $\kappa$-Minkowski spacetime coordinates and the spacetime
coordinates of a canonical spacetime. We observe that this ``cyclicity-inducing
measure'' is not invariant under $\kappa$-Poincar\'e transformations, but there
appears to be room for attempting to construct with such a measure a
$\kappa$-Poincar\'e invariant theory.

We start, in the next section, by reviewing some of the main properties of \kM\
noncommutative spacetime and of some possible choices of Weyl maps, which can
be used to introduce a correspondence between functions in \kM\ and commutative
functions of the $\lambda \rightarrow 0$ limit.
%%We also comment on the \kP\ symmetry
%transformations, as described by ``Weyl-map quantization''
%of the familiar translation, rotation and boost symmetry
%transformations.
In Section~III we introduce the problem of finding a cyclic action functional
and we construct explicitly a measure which ensures cyclicity. We also show
that the same choice of measure is compatible with all the Weyl maps considered
in Section~II. In Section~IV we examine the invariance properties of the action
functional. In Section~V we introduce the concept of a measure induced in \kM\
via a map which exists between $\kappa$-Minkowski spacetime coordinates and the
spacetime coordinates of a canonical spacetime, and we show that our
cyclicity-inducing
measure can be obtained in that way. Finally in Section~VI we comment on our
results and on the outlook of this research programme.

%%% ======================================================================

\section{$\kappa$-Minkowski star products}

\subsection{Preliminaries on star products and Weyl maps}
The natural framework for classical mechanics is a smooth manifold $M$ equipped
with a Poisson bracket $\{\,,\,\}$. In quantum mechanics the commutative
algebra $C^\infty(M)$ of smooth real-valued functions on the manifold is
replaced by a noncommutative $C^*$-algebra ${\mathcal{A}}$, and the Lie bracket
$[\,,\,]$, given by the commutator of two elements of the algebra, replaces the
classical Poisson bracket.

A \emph{quantization} is a one-to-one correspondence
$\Omega:C^\infty(M)\to  {\mathcal{A}}$ such that
\begin{displaymath}
[\Omega(f),\Omega(g)]=i\hbar\Omega(\{f,g\})+o(\hbar)\qquad\quad\forall\;f,g\in
C^\infty(M)
\end{displaymath}
This formulation~\cite{bayen1,bayen2} of the quantization problem, is
called \emph{deformation quantization}~\cite{gutt}, and the role
of deformation parameter is played by the Planck constant $\hbar$.

The \emph{Weyl map}~\cite{weyl} $\Omega$
establishes an isomorphism between
(a suitable subalgebra of) ${\mathcal{A}}$
and $C^\infty(M)$, if we equip the latter with a deformed product,
the star product ($*$-product or generalized Moyal-Weyl product~\cite{moy}),
implicitly defined by the equality
\be
\Omega(f*g)=\Omega(f)\cdot\Omega(g)
\label{stardef}
\ee

In the study of noncommutative spacetimes an approach based on the star
product
is widely used. It has proven very fruitful for the construction of theories in
canonical noncommutative spacetime, and it is expected that it should be also
useful in the analysis of Lie-algebra spacetimes, such as \kM . In
Ref.~\cite{Monaco} a construction of star products for a generic noncommutative
spacetime was presented, generalizing the Moyal-Weyl procedure with which the
star product of canonical spacetimes is obtained. An even more general
procedure, within an analysis that focused on  \kM , was discussed in
Ref.~\cite{alz02}.

We find useful to limit our analysis to three alternative choices of star
product for \kM , so that we can explore the possible Weyl-map dependence of
the results, while keeping a bearable level of complexity of
the discussion.

\subsection{Time-to-the-right star product}
The first star product we consider is the ``time-to-the-right''
star product, $*_R$, which has been widely used in the \kM\
literature
(see for
example~\cite{MajidRuegg,gacMajidgacMichele,KMLS,MO,KLM2}). The
corresponding Weyl map $\Omega_{R}$ can be introduced on the basis
of \be \Omega_R(f)=\frac{1}{(2\pi)^{2}}\int d^{4}k\tilde f(k)
e^{ik_j\x_j}e^{-ik_0\x_0} ~, \label{omegar} \ee where $\tilde{f}$
is the Fourier transform of the commutative function $f(x)$. This
map is ``time-to-the-right'' in the sense that $\Omega_R(e^{ik_j
x_j - ik_0 x_0}) = e^{ik_j\x_j}e^{-ik_0\x_0}$.

The star product $*_{R}$ associated with this Weyl map must of
course satisfy \be (f*_{R} g)(x)
=\Omega_{R}^{-1}\left(\Omega_{R}(f)\Omega_{R}(g)\right) ~. \ee
Using the identity \be
e^{ik_j\x_j}e^{-ik_0\x_0}\,e^{ip_j\x_j}e^{-ip_0\x_0}
=\Omega_R(e^{i\gamma^{\mu}_R(k,p)x_\mu}) \ee where \be
\gamma^\mu_R(k,p)= (k^0+p^0,k^j+e^{-\lambda k_0}p^j) \label{gamma}
\ee it is easy to see that the $*_R$ product for exponential
functions is\footnote{ Here we are using four-dimensional Greek
indices ($\mu,\nu=0,...,3$) and three-dimensional Latin indices
$(i,j=1,...,3)$. The short notations \emph{e.g.} $kx$ stand for
the contracted forms $k_\mu x^\mu$ with the $(-,+,+,+)$
signature.}: \bea e^{ik^\mu x_\mu}*_R e^{ip^\nu x_\nu}&=&
 \Omega_R^{-1}\left(\Omega_R(e^{ikx})\Omega_R(e^{ipx})\right)
= \Omega_R^{-1}\left(e^{ik_j\x_j}e^{-ik_0\x_0}\,
e^{ip_j\x_j}e^{-ip_0\x_0}\right)\nn\\
&=&\Omega_R^{-1}\left(\Omega_R(e^{i\gamma^\mu_R(k,p)x_\mu})\right)
=e^{i\gamma^\mu_R(k,p) x_\mu}
~. \label{expstarR}
\eea
Having specified the
$*_R$ product for exponential functions one of course
obtains the $*_R$ product for
generic functions through (\ref{omegar}).

\subsection{Time-Symmetrized Star Product}
As announced, we intend to explore the possible dependence of the
results on the choice of star product by considering a total of
three star products. The second in our list,
the ``time-symmetrized'' star product $*_T$, was first
introduced\footnote{Note that in earlier studies comparing
the time-symmetrized star product and the time-to-the-right
star product, such as the one of Ref.~\cite{aad03}, the index ``$S$''
rather than ``$T$'' was associated with the
time-symmetrized star product. Since we are here also considering
a fully symmetric star product (see later) we reserve the
index ``$S$'' for that choice.}
in Ref.~\cite{alz02}. It is based on the Weyl map $\Omega_{T}$,
introduced through \be \Omega_T(f)=\frac{1}{(2\pi)^{2}}\int
d^{4}k\, \tilde f(k)\,
e^{-i\frac{k_0\x^0}{2}}e^{ik_j\x_j}e^{-i\frac{k_0\x^0}{2}} \ee in
which again $\tilde{f}$ is the Fourier transform of the
commutative function $f(x)$.

The star product $*_T$ is of course such that
\be (f *_{T}
g)(x)=\Omega_{T}^{-1}\left(\Omega_{T}(f)\Omega_{T}(g)\right) ~,
\ee and in particular for exponentials one finds \bea
(e^{ikx}*_{T}
e^{ipx})&=&\Omega_{T}^{-1}\left(\Omega_{T}(e^{ikx})\Omega_{T}(e^{ipx})\right)\nn\\
&=&\Omega_{T}^{-1}\left(\Omega_{R}(e^{-ik_0x_0+i\vec{k}e^{-\lambda
k_0/2}\vec{x}})\Omega_{R}(e^{-ip_0x_0+i\vec{p}e^{-\lambda
p_0/2}\vec{x}})\right)\nn\\
&=&\Omega_{T}^{-1}\left(\Omega_{R}(e^{i\gamma^{\mu}_R(k_0,\vec{k}
e^{-\lambda k_0/2};p_0,\vec{p} e^{-\lambda p_0/2})x_\mu}\right)\nn\\
&=&\Omega_{T}^{-1}\left(\Omega_{T}(e^{i\gamma^{\mu}_R(k_0,\vec{k}
e^{\lambda p_0/2};p_0, \vec{p} e^{\lambda
k_0/2})x_\mu}\right)=e^{i\gamma^{\mu}_R(k_0,\vec{k} e^{\lambda
p_0/2};p_0, \vec{p} e^{\lambda k_0/2})x_\mu}\nn
 \eea
Thus one obtains
\be (e^{ikx}*_{T}
e^{ipx})=e^{i\gamma^{\mu}_T(k,p)x_\mu} \label{R-S}\ee where \be
\gamma^\mu_T(k,p)=\gamma^{\mu}_R(k_0,ke^{\lambda
p_0/2};p_0,pe^{\lambda k_0/2})=(k^0+p^0,k^je^{\lambda
p_0/2}+p^je^{-\lambda k_0/2}) ~, \ee which also exposes the simple
four-momenta transformation that relates the $\Omega_T$ Weyl map
and the $\Omega_R$ Weyl map: \be
\Omega_T(e^{ipx})=\Omega_R(e^{-ip_0x_0+i\vec{p}e^{-\lambda
p_0/2}\vec{x}}) ~. \ee

\subsection{Symmetric Star Product}
The third example of star product which we intend to consider, the
symmetric star product $*_{S}$, is the one adopted in
Refs.~\cite{wess0307149,lukieFT,Monaco}, and can be introduced
in association with the Weyl map $\Omega_{S}$:
$$ \Omega_{S}(f)
=\frac{1}{(2\pi)^{2}}\int d^4k\, \tilde f(k)\, e^{ik\x}
~.
$$

The symmetric star product $*_{S}$ is such that \be (f *_{S}
g)(x)=\Omega_{S}^{-1}\left(\Omega_{S}(f)\Omega_{S}(g)\right) ~.
\label{stardef1} \ee Using the Baker--Campbell--Hausdorff (BCH)
formula for the product of exponentials of noncommuting
quantities, one can easily verify the identity \be
e^{ik^\mu\x_\mu}e^{ip^\nu\x_\nu}=e^{i\gamma_S^\mu(k,p)\x_\mu}
\label{CBH} \ee where \bea
\gamma_S^\mu(k,p)=\left(k^0+p^0,\frac{\phi(k^0)e^{\lambda
p^0}k^j+\phi(p^0)p^j}{\phi(k^0+p^0)}\right) \eea with the function
$\phi(a)$ defined by $ \phi(a)=\frac{1}{a\lambda
}(e^{a\lambda}-1)$.

%%% ======================================================================

\section{Cyclic action functional for \kM }
In order to construct a field theory one
needs a linear functional, which will be used to introduce
an action functional. In the
commutative case the action functional is a map $I$
\be
I: C^\infty({\mathbb{R}}^4)\to {\mathbb{C}}
~,
\ee
and the familiar field theories in commutative spacetime
are based on the natural choice of action functional that
coincides with the ordinary integral:
\be
I(f)=\int {\de}^4x f(x)
\label{Icomm}
\ee
In the case of \kM\ the generalization of the action functional
will be given by a linear map ${\mathcal{I}}$:
\begin{displaymath}
{\mathcal{I}}: \kappa\mbox{-Minkowski} \to {\mathbb{C}}
\end{displaymath}
Since any element $\hat{f}(\x)$ of \kM\ can be written in terms of an
invertible Weyl map $\Omega$,
  $$ \hat{f}(\x)=\Omega(f(x)) ~,$$
a natural generalization of (\ref{Icomm}) would be
\be
{\mathcal{I}}\big(\Omega(f)\big)=\int {\de}^4x\, \mu(x)\,
\Omega^{-1}(\hat{f})(x)=\int {\de}^4x\, \mu(x)\,f(x)\label{Ikm}
\ee
where $\mu(x)$ is an appropriate integration measure.

In most of the works using the Weyl maps discussed in the
previous section the simple choice $\mu(x)=1$ is adopted:
\be
{\mathcal{I}}_1(\hat {f}) \equiv \int {\de}^4 \x \hat {f}(\x)
\equiv \int {\de}^4x f(x)
~,
\label{gacItriv}
\ee
where $\hat {f}(\x) = \Omega_R(f(x))$ in the cases in which
one adopts the time-to-the-right
star product, $\hat {f}(\x) = \Omega_T(f(x))$
for the time-symmetrized star product,
and $\hat {f}(\x) = \Omega_S(f(x))$ in the
cases with the symmetric star product.

The key objection to this choice of the trivial integration
measure, $\mu(x)=1$, is that the
resulting integral does not satisfy the cyclic property:
\be
{\mathcal{I}}_1(\hat{f}\hat{g}) \neq {\mathcal{I}}_1(\hat{g}\hat{f})
\ee
By renouncing to the cyclicity of the action functional
one looses a large number of familiar properties of a field theory,
and it is perhaps for this reason that the development of
field theories in \kM\ spacetime has not been very successful so far.

It is easy to see what should be required of the measure $\mu(x)$ in order to
achieve cyclicity. The requirement can be naturally expressed in terms of the
``star commutator'' $[f,g]_*\equiv f*g-g*f$. In fact, using the prescription
(\ref{Ikm}) and the star-product definition (\ref{stardef}), the action
functional for the product of two functions $\hat{f},\hat{g}$ of \kM\ is \bea
{\mathcal{I}}(\hat{f}\hat{g})&=&\int {\de}^4x\, \mu(x)\,
\Omega^{-1}(\hat{f}\hat{g})(x)=\int {\de}^4x\, \mu(x)\,
\Omega^{-1}(\Omega (f *_{\Omega} g))(x)\nn\\
&=&\int\,{\de}^4x\, \mu(x)\, f*_{\Omega}g \label{two} \eea and therefore the
cyclicity condition is \bea {\mathcal{I}}([\hat{f},\hat{g}])=\int\,{\de}^4x
\,\mu(x)\,[f,g]_*=0 \label{one} \eea A ``cyclicity-inducing measure'' should
satisfy the requirement (\ref{one}). Since the star commutator depends on the
particular choice of star product, we should contemplate the possibility that
the cyclicity-inducing measure, if it exists, may also depend on the choice of
star product.

In Ref.~\cite{wess0307149} it was claimed that a cyclicity-inducing measure
$\mu_S$ for the symmetric star product $*_S$ should exist, and that at first
order in $\lambda$ this measure should satisfy the requirement
$\vec{\nabla}(\vec{x}\,\mu_S(\vec{x}))=0$. We will give an explicit exact
(valid at all orders in $\lambda$) expression of a cyclicity-inducing measure,
and show that the same choice of measure is applicable to all three examples of
star product we are considering. In the later sections we will also show that,
while the measure can take a form that is clearly invariant under space
rotations, it is not possible to achieve the desired \kP\ invariance. And we
will observe that the existence of a cyclicity-inducing measure could be shown
straightforwardly exploiting the fact that there is a map between
$\kappa$-Minkowski spacetime coordinates and the spacetime coordinates of a
canonical spacetime.

\subsection{Cyclic action functional for the time-to-the-right star product}
We start by deriving the cyclicity-inducing
measure $\mu_R(x)$ for the case of the time-to-the-right star product.
The derivation is actually applicable to the general case
of a $D+1$-dimensional \kM\ spacetime,
and therefore  in the following we adopt conventions such
that vectors $\vec{v}$ are $D$-dimensional and latin index $j,k$
take values in $\{1,\ldots,D\}$.

The cyclicity condition (\ref{one}) must be satisfied for any $f$
and $g$, and, since
the functions $x^N=x_0^{n_0}...x_D^{n_D}$ form a basis,
it is sufficient for a cyclic action functional
to satisfy the following
conditions\footnote{We are assuming that the product $x^N*g(x)$ is integrable.}
\begin{equation}\label{eq:cond}
\int{\de}^{D+1}x\,\mu_R(x)[x_j^n, g(x)]_{*_R}=0 \qquad\quad
\int{\de}^{D+1}x\,\mu_R(x)[x_0^n, g(x)]_{*_R}=0
\end{equation}
for any natural number $n$.

The $*$-commutators can be analyzed considering the function
$g(x)=e^{ipx}$ and then extending the result to any function by
linearity, using the Fourier transform.

Noticing that
\bea
x_j*_Re^{ipx}&=&\lim_{q\to 0}(-i\partial_{q^j}e^{iqx}*_Re^{ipx})=
\lim_{q\to
0}(-i\partial_{q^j}e^{i\gamma_R(q,p) x})\nn\\
x_0*_Re^{ipx}&=&\lim_{q\to 0}(i\partial_{q^0}e^{iqx}*_Re^{ipx})= \lim_{q\to
0}(i\partial_{q^0}e^{i\gamma_R(q,p) x})\nn
\eea
one finds that the $[x_j^n,e^{ipx}]_{*_R}$
and $[x_0^n,e^{ipx}]_{*_R}$ star commutators can be written
in differential form as follows:
\bea
 {}[x_j^n,e^{ipx}]_{*_R}&=& (-i)^n\lim_{q\to
0}\partial_{q^j}^n
(e^{i\gamma_R(q,p)x}-e^{i \gamma_R(p,q) x})\nn\\
{}[x_0^n,e^{ipx}]_{*_R}&=& (i)^n\lim_{q\to 0}\partial_{q^0}^n
(e^{i\gamma_R(q,p)x}-e^{i \gamma_R(p,q) x})\nn
\eea
Using the explicit
expression (\ref{gamma}) of $\gamma_R(p,q)$ one then obtains
\bea
{}[x_j^n,e^{ipx}]_{*_R} &=&  [1-e^{-n\lambda p_0}](-i\partial_{p^j})^ne^{ipx} \nn\\
{}[x_0^n,e^{ipx}]_{*_R}&=& \{(-i\partial_{p_0}+i\lambda
p_j\partial_{p_j})^n-(-i\partial_{p_0})^n\}e^{ipx} \eea
and therefore the commutators in (\ref{eq:cond}) can be written as
\bea
 \ [x_j^n, g(x)]_{*_R} &=& x_j^n
(1-e^{in\lambda\partial_t}) g(x)
\label{eq3prime}\\
\ [x_0^n,g(x)]_{*_R} &=&\{(t+i\lambda x_j\partial_{x_j})^n-t^n\}g(x)
\label{eq3}
\eea
Substituting these expressions in (\ref{eq:cond}) and integrating by parts we
obtain the following differential equations for the cyclicity-inducing
measure $\mu_R(x)$
\bea
&&(1-e^{-in\lambda\partial_t})\mu_R(x)=0\label{eq1}\\
&&\left\{(t-i\lambda\vec{\nabla}\cdot\vec{x})^n-t^n\right\}\mu_R(\vec{x})=0
\qquad\forall\;n\geq 1 \label{eq2}
\eea
Eq.~(\ref{eq1}) implies that $\mu_R(\vec{x})$ does not depend on the
variable $t$, while (\ref{eq2}) gives rise to
the following series of equations:
\begin{eqnarray*}
\vec{\nabla}(\vec{x}\,\mu_R(\vec{x})) &=& 0 \\
\lambda^2\vec{\nabla}\vec{x}\,\vec{\nabla}\vec{x}\mu_R(\vec{x})
+2it\lambda\vec{\nabla}(\vec{x}\,\mu_R(\vec{x}))
&=& 0 \\
i\lambda^3\vec{\nabla}\vec{x}\,\vec{\nabla}\vec{x}\,\vec{\nabla}
\vec{x}\mu_R(\vec{x})-3\lambda^2t
\vec{\nabla}\vec{x}\,\vec{\nabla}\vec{x}\mu_R(\vec{x})-3i\lambda
t^2\vec{\nabla}(\vec{x}\,\mu_R(\vec{x})) &=& 0 \\
\ldots &=& 0
\end{eqnarray*}
A cyclicity-inducing measure will be obtained only
if all the equations of
the series are satisfied, and this happens when $\mu_R(\vec{x})$
is such that
\be
\vec{\nabla}(\vec{x}\,\mu_R(\vec{x}))=0\, .
\label{diffin}
\ee
This equation (\ref{diffin}) can be written in the form
\begin{equation}\label{omog}
\vec{x}\vec{\nabla}\mu_R(\vec{x})=-D\mu_R(\vec{x}) ~,
\end{equation}
% and the Euler theorem ensures that the solutions of this equation must be
% homogeneous functions with degree $-D$. Thus, (\ref{diffin}) is equivalent to
% \be \mu_R(a\vec{x})=a^{-D}\mu_R(\vec{x})\quad\forall\;a\in {\mathbb{R}},\;a\neq
% 0 \label{hom} \ee The only solution $\mu_R$ that preserves the desirable
% space-rotational invariance is, up to a multiplicative constant,
%which has no solutions that are
%continuous and derivable in all $\R^D$. We can however solve it
%in $\R^D-\{0\}$.
which, if we search for a measure that preserves space-rotational
invariance ($\mu=\mu(|\vec{x}|))$, is equivalent to
\begin{displaymath}
|\vec{x}|\frac{\partial}{\partial |\vec{x}|}\mu_R(|\vec{x}|)=-D\mu_R(|\vec{x}|)
~.
\end{displaymath}
The only solution is, up to a multiplicative constant,
\begin{equation}\label{eq:sol}
\mu_R(\vec{x})=|\vec{x}|^{-D} ~ .
\end{equation}
We therefore can formulate an action functional which enjoys the sought
cyclicity\footnote{Clearly this choice of measure
is only acceptable when the function $f(t,\vec{x})$ is
such that $f(t,0)=0$. The reader can however easily verify that
the measure can be extended to the case $f(t,0)\neq 0$. We shall not
dwell on this point since, for independent reasons, one naturally
considers~\cite{wess0307149}
for theories in \kM\ the choice of ``Lagrangian densities''
that vanish for $\vec{x}=0$.
We shall also comment on this point in our closing remarks.}:
\begin{equation}
\label{goal}
{{\mathcal{I}}(\Omega_R(f))}=\int\frac{1}{|\vec{x}|^D}f(x)\,\de^{D+1}x
\end{equation}

\subsection{Cyclic action functional for the time-symmetrized star product}
We now show that the same choice of measure that
induces cyclicity of the action functional for the
time-to-the-right star product also induces cyclicity of
the action functional for the time-symmetrized star product. We
proceed searching for a cyclicity-inducing $\mu_T(x)$ and then
verify that $\mu_T(x)=\mu_R(x)$.

Cyclicity requires a $\mu_T(x)$ such that\footnote{We are
going back to focusing
on the case $D=3$, but, as in the previous star-product case, the
discussion can be easily generalized to an arbitrary number of
dimensions.} \be \int {\de}^4x\,\mu_T(x)[f(x),g(x)]_{*_{T}}=0 \ee
{\it i.e.}
\begin{displaymath}
\int {\de}^4x\,\mu_T(x)[x_j^n, g(x)]_{*_T}=0 \qquad\quad
\int{\de}^4x\,\mu_T(x)[x_0^n, g(x)]_{*_T}=0
\end{displaymath}
for all the integer $n$.

As already stressed in (\ref{R-S}), the time-to-the-right Weyl map
$\Omega_R$ and the time-symmetrized Weyl map $\Omega_T$ are
connected in the following way: \be
\Omega_R(e^{ipx})=\Omega_{T}(e^{ip'x}) \label{change} \ee where
$p'=(p_0,e^{\lambda p_0/2}p_j)$.

This can be used to show that from \be [x_j^n, e^{ipx}]_{*_R}=
(1-e^{-n\lambda p_0})(-i\partial_{p^j})^ne^{ipx} \ee it follows
that \bea [x_j^n, e^{ipx}]_{*_{T}}&=&(1-e^{-n\lambda
p_0})e^{n\lambda p_0/2}(- i\partial_{p_j})^n e^{ipx}\nn \eea and
therefore \be [x_j^n, g(x)]_{*_{T}}=x_j^n(1-e^{in\lambda
\partial_t})e^{-in\lambda \partial_t/2} g(x)
\ee
The form of this equation differs slightly from the corresponding
equation obtained for the time-to-the-right star product,
but the associated requirement for $\mu_T$
\be
e^{-in\lambda
\partial_t/2}(1-e^{-in\lambda\partial_t})\mu_T(x)=0
\ee
still leads to the conclusion that the
cyclicity-inducing
measure should not depend on $t$.

Once this $t$-independence is taken into account, in order to
ensure cyclicity of the integral we are left with the task of
enforcing \be \int {\de}^4x \mu({\vec{x}}) [x_0^n ,g(x)]_{*_T}=0
\ee or equivalently (using Fourier series for $x_0^n,g(x)$)
\be (-i)^n\int {\de}^4x\,\mu({\vec{x}})\, {\de}^4k\,{\de}^4p\,
\left(\partial_{k_0}^n\delta^{(4)}(k)\right) \tilde{g}(p) [e^{ikx}
,e^{ipx}]_{*_T}=0 \label{jjj} \ee
Using
\be \gamma_T^{\mu}(k,p)=(k^0+p^0,e^{\lambda
p^0/2}k^j+e^{-\lambda k^0/2}p^j) \ee the condition (\ref{jjj}) can
be rewritten as
\be (-i)^n \int {\de}^4x \mu({\vec{x}})\, {\de}^4k\,{\de}^4p\,
\left(\partial_{k_0}^n\delta^{(4)}(k)\right)\tilde{g}(p)
[e^{i\gamma_T(k,p)x}-e^{i\gamma_T(p,k)x}]=0\ee
Then, upon integration in $dt$ and in $d^4k$, one obtains
\bea
 &&\int {\de}\vec{x} \mu({\vec{x}})\,{\de}^4p\, \left(\partial_{k_0}^n\delta(k_0)\right)
\big|_{k_0=-p_0}\tilde{g}(p)
e^{ip^je^{\lambda p_0/2}x_j}=~~~~~~~~~~~\nn\\
&& ~~~~~~~~~=\int {\de}\vec{x} \mu({\vec{x}})\,{\de}^4p\,
\left(\partial_{k_0}^n\delta(k_0)\right) \big |_{k_0=-p_0}\tilde{g}(p)
e^{ip_je^{-\lambda p_0/2}x_j} \nn \eea
Performing the substitution $x^j\to e^{-\lambda p_0/2}x^j$ on the left-hand
side, and the substitution $x^j\to e^{\lambda p_0/2}x^j$ on the right-hand
side, one then obtains
\bea &&\int {\de}\vec{x}\,{\de}^4p\, \qs e^{-3\lambda p_0/2} \mu(e^{-\lambda
p_0/2}\vec{x})-e^{3\lambda p_0/2}\mu({e^{\lambda p_0/2}\vec{x}})\qd
(\partial_{k_0}^n\delta(k_0))|_{k_0=-p_0}\tilde{g}(p) e^{ip_jx_j}=0\nn \eea
which can be satisfied if and only if
\be \mu_T(a\vec{x})=a^{-3}\mu_T(\vec{x})\label{e1}\ee
%for all $a=e^{\lambda p_0}\in\R^+$.
for all $a \in \R^+$.

If we search for a measure that preserves space-rotational invariance, the only
solution is, up to a multiplicative constant,
\begin{displaymath}
\mu_R(\vec{x})=|\vec{x}|^{-3} ~ .
\end{displaymath}
The form of $\mu_R(\vec{x})$ discussed earlier
is therefore also an acceptable cyclicity-inducing choice
of $\mu_T(\vec{x})$.

\subsection{Cyclic action functional for the symmetric star product}
Next we turn to the case of the symmetric star product $*_S$.
As mentioned, this is the case considered in Ref.~\cite{wess0307149},
where the first remarks on the possibility of a ciclicity-inducing
measure were made.

The cyclicity condition (\ref{one}) for the symmetric star product
\be \int {\de}^4x\,\mu_S(x)[f(x),g(x)]_{*_{S}}=0 \ee can be
satisfied by imposing
\begin{displaymath}
\int{\de}^4x\,\mu_S(x)[x_j^n, g(x)]_{*_S}=0 \qquad\quad
\int{\de}^4x\,\mu_S(x)[x_0^n, g(x)]_{*_S}=0
\end{displaymath}
for all the integers $n$.

The analysis proceeds in close analogy to the case of the
time-symmetrized star product considered in the previous
subsection. In particular, for the first commutator one finds that
\bea [x_j^n,g(x)]_{*_{S}}&=&(1-e^{-n\lambda
p_0})\phi^{-n}(p_0)(-i\partial_{p_j})^n g(x)\nn\\
&=&x_j^n(1-e^{in\lambda
\partial_t})\phi^{-n}(-i\partial_t) g(x)
~,
\eea
and the corresponding condition for $\mu_S(x)$,
\be
\phi^{-n}(-i\partial_t)(1-e^{-in\lambda\partial_t})\mu_S(x)=0
\ee
implies that $\mu_S(x)$ does not depend on $t$.

The residual requirement for  $\mu_S(x)$ can be written in the form \be \int
{\de}^4x \mu_S({\vec{x}}) [x_0^n ,g(x)]_{*_S}=0 ~, \label{Fcond} \ee which is
equivalent to
\bea && \int {\de}^4p\, {\de}^3x\, \mu_S({\vec{x}})
\tilde{g}(p)[\partial_{k_0}^n\delta(k_0)]_{k_0=-p_0} e^{i
p^j \phi(p^0)x_j}= ~~~~~~~~~~~~~~~~ \nn\\
&& ~~~~~~~~~~~~~~~  = \int {\de}^3k\,{\de}^3x\, \mu_S({\vec{x}})
[\partial_{k_0}^n\delta(k_0)]_{k_0=-p_0} e^{ip^je^{\lambda p_0}\phi(p_0)x_j}
\nn ~, \eea
where we have taken into account that $\phi(-p_0)=e^{\lambda p_0}\phi(p_0)$ and
$\phi(0)=1$.

Therefore $\mu_S({x})$ must be such that
\bea && \int {\de}^4p\, {\de}^3x\, \phi^{-3}(p_0) \qs
\mu_S(\phi^{-1}(p_0){\vec{x}})-e^{-3\lambda
p_0}\mu_S(\phi^{-1}(p_0)e^{-\lambda p_0}\vec{x})\qd
\tilde{g}(p)\,\partial_{k_0}^n\delta(k_0)|_{k_0=-p_0}
e^{ip^jx_j}=0 \nn \eea
and this condition can be satisfied if and only if \be
\mu_S(a\vec{x})=a^{-3}\mu_S(\vec{x})\label{e2} \ee for all $a \in
\R^+$. This is the same requirement encountered already for the
cases of the time-symmetrized star products. If we require rotational
invariance, for all three examples ($*_R$, $*_T$ and $*_S$) of
star products the integration measure is (\ref{eq:sol}).

%%% ======================================================================
\section{Symmetry analysis}
\subsection{Symmetry operators in the commutative case}
One of the most studied aspects of \kM\ is the possibility that it might
be invariant under an Hopf-algebra of symmetries known as \kP.
This is a
rather technical subject which has been discussed extensively
in the literature
(see, {\it e.g.},
Refs.~\cite{lukieFT,gacMajidgacMichele,aad03,MajidRuegg,lukieAnnPhys,jurekKMKP}).
Here it is sufficient for us to introduce the \kP\ Hopf-algebra transformations
in an elementary way. For definiteness we focus on the case of the
time-to-the-right star product (but analogous results hold for the other
choices of star product which we are considering), and we also find sufficient
to consider the 1+1-dimensional \kM\ spacetime.

Just as a way to fix notations and terminology,
in this subsection we review briefly
the situation in the commutative-spacetime case.
If $x\to x'$ is an element of a fixed transformation group $G$,
a function $f$ is
called \emph{scalar} for $G$ if it transforms as \be f\to f'\; ,\quad
f'(x')=f(x)\label{scalar} \ee Given a scalar Lagrangian (that for us is simply
a scalar function), in order to construct a $G$-invariant action, we need an
invariant integral, {\it i.e.} an integral satisfying
\begin{displaymath}
\int f'(x)=\int f(x)
\end{displaymath}
for all transformations of $G$ and for all the scalar functions $f$.

If $f$ is a function of the classical Minkowski spacetime coordinates,
and one adopts
the standard integral $\int {\de}^{D+1}x \,f(x)$,
then the relevant symmetry group is the Poincar\'{e}
group.

For an integral with a non-trivial measure
one can observe that
\begin{displaymath} \int f'=\int
{\de}^{D+1}x\;\mu(x)\, f'(x)=\int {\de}^{D+1}x'\;\mu(x') \,f'(x')
\end{displaymath}
and that, using (\ref{scalar}),
\begin{displaymath}
\int f'=\int {\de}^{D+1}x'\;\mu(x') \,f(x)
\end{displaymath}
Therefore the integral is
invariant if and only if $\mu(x)\de^{D+1}x$ is invariant.

For a generic $G$, if
\begin{displaymath}
x\to x'=x+A(x)\;,
\end{displaymath}
with $A:{\mathbb{R}}^4\to {\mathbb{R}}^4$ smooth, is a transformation of
coordinates, the operators $A^\nu(x)\partial_\nu$
form a Lie
algebra and their exponentiation $e^{iA^\nu(x)\partial_\nu}$
give us a Lie group which describes
the transformation rule of scalar functions
\begin{displaymath}
f'(x)=e^{-iA^\nu(x)\partial_\nu}f(x)\quad\iff\quad f'(x+A(x))=f(x)
\end{displaymath}
For $T=A^\nu(x)\partial_\nu$ a generator of the group, if we denote
\begin{equation}\label{trivial}
\Delta T=T\otimes 1+1\otimes T
\end{equation}
we can rewrite the Leibniz rule as
\begin{equation}\label{eq:cdot}
T(f\cdot g)=(T_{(1)}f)(T_{(2)}g)
\end{equation}
where $\Delta T=T_{(1)}\otimes T_{(2)}$ is the Sweedler notation. The operation
$\Delta$ (the \emph{coproduct}) is extended to all the universal enveloping
algebra of $G$ by the request that it should be
an algebra-morphism and gives a structure
of (trivial) Hopf-algebra.

A coproduct such as (\ref{trivial}) is ``called trivial''.
When we consider a
noncommutative algebra of functions (as in the case of the study
of theories in noncommutative spacetimes),
in general the equation (\ref{eq:cdot}) cannot be
satisfied by a trivial coproduct, but still a Hopf-algebra structure
can emerge.

It appears~\cite{aad03} that a description of symmetries
in terms of Hopf algebras is appropriate both
for commutative and for noncommutative spacetimes
(but when the spacetime is commutative the co-algebra sector
is trivial, and a description in terms of a Lie algebra would suffice).

\subsection{\kP\ and the symmetries of the non-cyclic
action functional}
The analysis of the symmetries of the action functional
in the \kM\ noncommutative spacetime can be set up in complete
analogy with the commutative case discussed in the previous
subsection.
In \kM , analyzed in terms of the $\Omega_R$ time-to-the-right Weyl map,
the action functional
\begin{displaymath}
{\mathcal{I}}(\Omega_R(f))=\int {\de x\de t}\, \mu(x,t)\,f(x,t) ~,
\end{displaymath}
where $f(x)$ is a scalar function, is invariant
under a transformation $T$ if and only if
\be
\int {\de}x\,{\de}t\mu(x)\;f'(x)=\int {\de}x\,{\de}t\mu(x)\;f(x)\;.
\ee

For each generator $T$ of the symmetry
transformations, we define
\begin{equation}\label{eq:Tstar}
T\big[\Omega_R(f)\Omega_R(g)\big]=[T_{(1)}\Omega_R(f)][T_{(2)}\Omega_R(g)]\;\;,
\end{equation}
which is the analogue of (\ref{eq:cdot}).

In order to have a genuine symmetry algebra, an algebra that can be used
to describe all aspects of the symmetries of the action functional,
it is necessary~\cite{aad03} that $T_{(1)}$ and $T_{(2)}$
involve only operators of the algebra.
For the commutative case considered in the previous subsection
this request is automatically satisfied, but (as it will become apparent
later on in our analysis) the noncommutativity of
the spacetime coordinates can change the situation significantly.
When $T_{(1)}$ and $T_{(2)}$ (for all $T$ in the symmetry algebra)
involve only operators of the algebra one inevitably obtains~\cite{aad03}
the structure of a Hopf algebra of symmetries. We will
therefore
adopt terminology such that the requirement that the would-be symmetry
generators close a Hopf algebra is identified with the description
of a symmetry algebra.

As mentioned, a large literature has been devoted to the
possibility that field theories in \kM\ might
provide a realization of \kP\ Hopf-algebra symmetries.
In order to see how \kP\ can naturally emerge let us start
by introducing in \kM\ translation and rotation transformations
naturally obtained by acting with the Weyl map on the
corresponding transformations of the commutative limit:
\begin{equation}\label{traslone}
P_\mu \Omega_R(f)=\Omega_R(-i\partial_\mu f)
\end{equation}
\begin{equation}\label{rotone}
M_j \Omega_R(f)=\Omega_R(i\epsilon_{jkl}x_k\partial_lf)
\end{equation}
One can easily verify~\cite{aad03}, imposing (\ref{eq:Tstar}),
that $P_0$ and $M_j$ have trivial coproduct, while
\begin{displaymath}
\Delta P_j=P_j\otimes 1+e^{-\lambda P_0}\otimes P_j
\end{displaymath}

A key point in the analysis of \kM\ is the observation that these descriptions
of translations and space rotations, in terms of Weyl-map quantizations of the
corresponding classical transformations, are incompatible with the description
of boost transformations given by Weyl-map quantization
\begin{equation}\label{boostone}
\tilde{N}_j \Omega_R(f)=\Omega_R(i[x_0\partial_j-x_j\partial_0]f)
\end{equation}
In fact acting with (\ref{boostone}) on the product of two element of \kM\
implicitly requires~\cite{aad03} a description in terms of an operator which is
external to the algebra. Consistency with the Hopf-algebra requirements leads
to the introduction~\cite{aad03,MajidRuegg,lukieAnnPhys} of the ``deformed
boost action''
\begin{equation}\label{boostdef}
N_j=it\partial_j-x_j\left(\frac{1-e^{2i\lambda\partial_t}}{2\lambda}
-\frac{\lambda}{2}\nabla^2\right)-i\lambda x_l\partial_l\partial_j
\end{equation}
The elements $P_\mu$, $M_j$ and $N_j$ generate the \kP\ Hopf-algebra
(in the Majid-Ruegg basis~\cite{MajidRuegg}).

As a first step in the exploration of the possibility of a \kP\ invariant
field theory in \kM , we can ask if \kP\ can be realized as symmetry group
of the action functional. This is basically the reason that motivated the
choice of measure $\mu =1$ in most of the early works on \kM. In fact,
the choice $\mu=1$ leads to a \kP\ invariant action functional.
For our purposes here it is
sufficient to revisit this result considering only the case in which
one adopts the time-to-the-right star product, and focusing on the
case of a 1+1-dimensional spacetime.

In the noncommutative case
the symmetry analysis involves several new elements of complexity,
especially in light of the fact that, since the
coproduct is deformed, it is no longer true that $f'(x')=f(x)$,
with $x'=e^{-iaT}x$,
if $T$ is a generator of a
one-parameter transformation and $f'=e^{iaT}f$.
This equality holds if and only if $T$ has a
trivial coproduct. In fact, only when $\Delta T$ is trivial,
the invariance condition
\begin{equation}\label{eq:sym}
\int f'(x,t)\mu(x,t)\de x\de t=\int f(x,t)\mu(x,t)\de x\de t
\end{equation}
is equivalent\footnote{In particular, in our \kP\ context
the space-rotation generators $M_j$ have a trivial coproduct,
so it
makes sense to state that a rotationally-invariant measure $\mu(|\vec{x}|)$
gives a rotationally-invariant integral.
(Of course, this remark is relevant when working with more than one
space dimension, since
in 1+1 dimensions there are no space rotations).}
to the invariance of the integration
measure: $\mu(x',t')\de x'\de t'=\mu(x,t)\de x\de t$.

It is actually easy to see, in the analysis of
the symmetries of the action functional
\begin{displaymath}
\int f(x,t)\de x\de t
\end{displaymath}
in the sense of Eq.~(\ref{eq:sym}),
that time translations and space translations, generated by
$P_\mu$, are symmetries; indeed
\begin{displaymath}
\int \{e^{iaP}f\}(x,t)\de x\de t=\int f(x,t)\de x\de t
\end{displaymath}
In other words, the infinitesimal variation is zero:
\begin{displaymath}
\int (P_\mu f)(x,t)\de x\de t=0
\end{displaymath}
It is also easy to verify that the deformed boost is a symmetry,
indeed integration by parts gives
\begin{displaymath}
\int (N_j f)(x,t)\de x\de t=0
\end{displaymath}

The requirement that the generators of a genuine symmetry algebra
should close on to a Hopf-algebra
is also satisfied since $P_\mu$ and $N_j$
are the generators of the well-known \kP\ Hopf algebra.

\subsection{Symmetries of the cyclic action functional}
The fact that the simple
choice of measure $\mu =1$ leads to a \kP\ invariant action functional
has geerated a strong interest in the literature. Some of the
reasons for this interest originate from a
possible ``\kP\ phenomenology''~\cite{polonpap,gianluFranc}
and some other reasons of interest
reside deep in the rather rich mathematical
structures involved.
Here we just want to stress one obvious attractive aspect
of the \kP\ symmetries: in the commutative $\lambda \rightarrow 0$
limit they reduce to standard (classical, Lie-algebra)
Poincar\'e symmetries. Therefore the \kP\ invariance is fully compatible
with the fact that our low-energy experiments (involving distance
scales much larger than $\lambda$) are all consistent, within
their available accuracy, with classical Poincar\'e invariance.

However, as stressed above, the fact that with the measure $\mu =1$
the action functional is ``non-cyclic'' leads
to several problems.
With our cyclic action functional
these problems would be avoided, but we intend to observe in this
subsection that the covariance properties
of the cyclic action functional appear to be somewhat peculiar.
We can establish this point already by working again
with $1+1$-dimensional \kM , where the cyclic action
functional takes the form
\begin{displaymath}
{\mathcal{I}}(\Omega_R(f))=\int f(x,t)\,\frac{\de x}{|x|}\,\de t
\end{displaymath}

The Weyl-map quantization
of the time translation is clearly a symmetry, and is generated
by $P_0=-i\partial_t$. Indeed, it keeps invariant the element
\begin{displaymath}
\mu(x)\de x\de t=|x|^{-1}\de x\de t
\end{displaymath}
and, since it has a trivial coproduct, it leaves invariant
also the integral, in
the sense of Eq.~(\ref{eq:sym}).

On the other hand, it is equally easy to verify that
the space translation and
the boost (both the classical boost and the \kP\ deformed
boost) are not symmetries of the cyclic action functional.
These two symmetries are replaced by two other, possibly unwanted,
symmetries.
One is the Weyl-map quantization
of an $x$-dilatation, with generator $D=-ix\partial_x$.
The finite transformation obtained with $D$
is $f\to f'=e^{-i(\log a)D}f$, with $a\in {\mathbb{R}}^+$.
In order to establish the form of $\Delta D$, the coproduct of $D$,
it is sufficient to
take $f(x,t)=e^{ipx}$ and $g(x,t)=e^{iqx}$ with $p$ and $q$ arbitrary, so
that $(f*g)(x,t)=e^{i\gamma_R(p,q)x}$, where $\gamma_R(p,q)$ is
the one given in Eq.~(\ref{gamma}). One can then easily check that
\begin{displaymath}
\partial_x(f*g)=(\partial_x f)*g+(e^{-\lambda P_0}f)*(\partial_x g)
\end{displaymath}
Moreover, $x*f=xf$ and therefore (by associativity of the $*$-product)
$(xf)*g=x*f*g=x(f*g)$, while $e^{ip_0t}*x=x*e^{ip_0t}e^{-\lambda p_0}$.
Thus
\begin{displaymath}
f*(xg)=f*x*g=e^{-\lambda p_0}x*f*g=x[(e^{-\lambda P_0}f)*g]
\end{displaymath}
and from this one concludes that
\begin{displaymath}
x\partial_x(f*g)=x[(\partial_x f)*g]+x[(e^{-\lambda P_0}f)*(\partial_x g)]
=(x\partial_x f)*g+f*(x\partial_x g)
~,
\end{displaymath}
 {\it i.e.} $\Delta D=\Delta(x\partial_x)$ is trivial.

The fact that
$D$ has a trivial coproduct implies that $f$ is
scalar under $D$ when $f'(x,t)=f(ax,t)$.
Since\footnote{One could formally consider
also $a<0$ but that would not be a genuine dilatation.
It combines a dilatation with an inversion, $x \rightarrow -x$.
The cases with $a<0$ will be obtained
combining an $a>0$ dilatation and a space-rotation
such that $x \rightarrow -x$.}
 $a>0$, the fact that $D$ is a symmetry follows from
the corresponding
invariance of the integration measure: $\de x/|x|=\de (ax)/|ax|$.

While the form of the cyclicity-inducing measure
immediately suggests that dilatations and time translations
should be symmetries,
the identification of the third symmetry, which
we denote by $K$,
requires more work.
However, writing $K$ as a formal series in the coordinates and derivatives
one can straightforwardly (but tediously)
reconstruct the form of this third symmetry, obtaining the result
\begin{displaymath}
K=-itx\partial_x+\lambda\log|x|\,\left(\frac{1-e^{2i\lambda
\partial_t}}{2}+\frac{1}{2}(x\partial_x)^2\right)
~.
\end{displaymath}
The fact that this $K$
generates a symmetry of the cyclic-integral, {\it i.e.}
${\mathcal{I}}(\Omega(e^{i\xi K}\,f))={\mathcal{I}}(\Omega(f))$
for all $\xi$
(or equivalently ${\mathcal{I}}(\Omega(K f))=0$),
is quickly verified observing that
integration by part gives
\begin{displaymath}
\int (K f)(x,t)\,\frac{\de x}{|x|}\,\de t=\int f(x,t)\left[
\partial_x\frac{itx}{|x|}+\frac{\lambda}{2}\left((1
-e^{-2i\lambda\partial_t})\frac{\log|x|}{|x|}
+\partial_x^2\frac{x^2\log|x|}{|x|}
-\partial_x\frac{x\log|x|}{|x|}\right)\right]\de
x\,\de t = 0
~.
\end{displaymath}

We do have three symmetry candidates, $P_0$, $D$ and $K$;
however, these are not clearly the symmetries we were looking for,
since in the commutative $\lambda \rightarrow 0$ limit
they do not recover the classical Poincar\'e algebra.
On the other hand this should be expected since
in the $\lambda \rightarrow 0$ limit the cyclicity-inducing
measure does not reduce to the Poincar\'e invariant measure $\mu =1$.

Also alarming is the fact that the three
operators $P_0$, $D$ and $K$ do not generate e genuine
symmetry algebra, not in the sense needed when dealing with a noncommutative
spacetime. In fact,
the expression of
their co-products requires operators external to the
triad $P_0$, $D$ and $K$, so $P_0$, $D$ and $K$ do not generate
a Hopf algebra.
This is an automatic consequence of the fact that
both $P_0$ and $D$ have trivial coproduct, while
the commutator $[K,D]$ is nonlinear:
\begin{displaymath}
[K,D]=i \lambda \left(\frac{1-e^{-2\lambda
P_0}}{2}\right)
- i \frac{\lambda}{2}D^2
\end{displaymath}
In fact, the Hopf-algebra axioms imply that $\Delta K$ is of the form
\begin{displaymath}
\Delta K=K\otimes a(P_0,D)+b(P_0,D)\otimes K
\end{displaymath}
with $a$ and $b$ to be determined.
But the condition $[\Delta K,\Delta P_0]=\Delta[K,P_0]=i\lambda\Delta D$
(where we used the observation that $[K,P_0]=i\lambda D$)
implies $a=b=1$, and for $a=b=1$ the
condition $[\Delta K,\Delta D] = \Delta[K,D]$
is not satisfied
\begin{displaymath}
[\Delta K,\Delta D]-\Delta[K,D]=\textrm{$\frac{i}{2\lambda^2}$}(1-e^{-2\lambda
P_0})\otimes (1-e^{-2\lambda P_0})-iD\otimes D\neq 0
\end{displaymath}

%%% ======================================================================
\section{Cyclicity-inducing measure for \kM\ derived from
canonical-spacetime action functionals}
In the previous sections
we have shown  that
a cyclicity-inducing measure for \kM\
can be obtained explicitly using an analysis which only
relies on properties of \kM\ itself.
In this section we intend to show that the same results
can be obtained even more easily
by exploiting the properties
of maps which allow to formulate
the $\kappa$-Minkowski spacetime coordinates in terms of
the spacetime coordinates of a canonical spacetime.

This may prove also valuable as a starting point for future
use of our results on the action functional for the
formulation of field theories in \kM .
In fact, canonical spacetimes are rather well understood
and perhaps could be exploited as a starting point
for further analysis of \kM .

There are actually two ways to obtain a cyclicity-inducing
measure for \kM\ starting from canonical spacetimes: one
which is based on a direct one-to-one description
of \kM\ coordinates in terms of canonical coordinates
and one which is based on a description of
the \kM\ coordinates in terms of a higher-dimensional
canonical spacetime.

\subsection{A procedure involving simply a Jacobian}
Let us start by observing that
the relations
\begin{displaymath}
x_j=e^{q_j}
\end{displaymath}
give us an isomorphism of the first quadrant of \kM\
with ${\mathbb{R}}^{D+1}$ equipped with commutation relations:
\begin{displaymath}
[q_j,q_k]_*=0\qquad\quad [q_j,t]_*=i\lambda
\end{displaymath}
For any given star product on \kM\
there is a corresponding star product on the spacetime
with $(q,t)$ coordinates.
Since the $(q,t)$ algebra is canonical,
the action functional $\int F(\vec{q},t)\de^Dq\,\de t$
is cyclic for any choice of star product.

We seek a space-rotations invariant action functional,
so, for a $D+1$-dimensional \kM , we can
divide the $D$-dimensional space in $2^D$ quadrants.
In each quadrant all the coordinates $x_j$ have a
fixed sign. If we determine ${\mathcal{I}}(\hat{f})$ for $f$ zero
everywhere except when all $x_j>0$, then by rotational invariance
we also determine ${\mathcal{I}}(\hat{f})$ for a function different
from zero in any other quadrant.
We can therefore focus on the first quadrant:
\begin{displaymath}
{\mathcal{I}}(\hat{f})=\int_{x_j\geq 0\;\forall\;j}
f(x)\,\mu(|\vec{x}|)\,\de^{D+1}x\qquad,\; f(0,t)=0
\end{displaymath}

Using the points made above about the map $x_j \rightarrow e^{q_j}$
we can describe a function $f(x_1,\ldots,x_D,t)$
of the \kM\ coordinates
as a function $F(\vec{q},t)=f(e^{q_1},\ldots,e^{q_D},t)$
of the $(q,t)$ coordinates.
Then the classical integral in the $(q,t)$ coordinates
corresponds to a cyclic action functional in \kM:
\begin{displaymath}
\int F(\vec{q},t)\de^Dq\,\de t=\int_{x_j\geq 0\;\forall\;j}
f(x)\,J(\vec{x})\,\de^{D+1}x
\end{displaymath}
where
\begin{displaymath}
J(\vec{x})=\frac{1}{|x_1|\ldots |x_D|}
\end{displaymath}
is the Jacobian of the transformation.
Note
that $J(\vec{x})$ is a particular solution of (\ref{omog}), as
one should expect.

This particular cyclicity-inducing measure is not
space-rotation invariant.
One can obtain
a space-rotation invariant measure for \kM\ starting from a
cyclic integral in the canonical spacetime that is different from
the classical one.
Formulating the star product as a
pseudo-differential operator, and integrating by parts,
one finds that an integration measure $\eta(\vec{q},t)$ gives a cyclic
action functional if it satisfies:
\begin{displaymath}
(\partial_{q_1}+\ldots+\partial_{q_D})\eta(\vec{q},t)
=0\qquad\quad\partial_t\eta(\vec{q},t)=0
~.
\end{displaymath}
It should be stressed that this
is not simply the equation (\ref{omog}) rewritten in the new
coordinates.
However,
if we choose the particular solution
\begin{displaymath}
\eta(\vec{q})=\exp\left(q_1+q_2+\ldots+q_D
-\frac{D}{2}\log[e^{2q_1}+\ldots+e^{2q_D}]\right)
\end{displaymath}
we do obtain our favoured, cyclicity-inducing and space-rotation
invariant, measure for \kM :
\begin{displaymath}
\mu(\vec{x})=J(x)\cdot\exp\left(\log|x_1|
+\ldots+\log|x_D|-\frac{D}{2}\log|\vec{x}|^2
\right)=J(x)\exp\left(\log\frac{|x_1|\ldots|x_D|}{|\vec{x}|^D}\right)
=\frac{1}{|\vec{x}|^D}
\end{displaymath}

%%% ======================================================================
\subsection{Cyclic action functionals by reduction}
It is possible to view~\cite{selene} a $D+1$-dimensional
($D \geq 1$) noncommutative spacetime ${\mathcal{M}}$ as a subspace of a $2D$
dimensional symplectic phase space ${\mathcal{C}}$, endowed with the
usual Poisson Bracket, and a Moyal star product.
A star product on ${\mathcal{M}}$ can be
defined by first lifting the functions from
the $(D+1)$-dimensional space ${\mathcal{M}}$
to functions on the $2D$-dimensional phase space ${\mathcal{C}}$,
then multiplying them in such a phase
space using the Moyal product and finally projecting back to the
original space ${\mathcal{M}}$.
The lift consists in establishing a (generalized Jordan-Schwinger)
map between the coordinates of the space ${\mathcal{M}}$ and
the coordinates of the phase space ${\mathcal{C}}$.

Examples of such maps for \kM\ have been known for some time (see,
{\it e.g.}, Refs.~\cite{gacMajidgacMichele,alz02}) and they have recently been
used~\cite{alz02} to introduce star products in \kM . In this
section we consider the case in which ${\mathcal{C}}$ is equipped
with $2D$ canonical coordinates $(x_j,p_j)$ and the
Groenewold~\cite{groen} star product
\begin{equation}\label{tre}
(F*G)(x,p)=F(x,p)e^{i(\overleftarrow{\partial}_{x_j}
\overrightarrow{\partial}_{p_j}- \overleftarrow{\partial}_{p_j}
\overrightarrow{\partial}_{x_j})}G(x,p)
\end{equation}
The commutation rules are those of the Heisenberg algebra
\begin{displaymath}
[x_j,p_k]_*=i\delta_{jk}\qquad [x_j,x_k]_*=[p_j,p_k]_*=0
\end{displaymath}
Introducing the map
\begin{displaymath}
x_0=\lambda\vec{x}\vec{p}
\end{displaymath}
one can easily to verify that the $x_\mu$ satisfy the \kM\
commutation relations
\begin{displaymath}
[x_j,x_0]_*=\lambda[x_j,\vec{x}\vec{p}]=i\lambda x_j
\end{displaymath}
The star product in \kM\
induced by this map is different~\cite{alz02} from
the three star products that we have so far considered.

In the phase
space ${\mathcal{C}}$, the standard integral $\int
F(\vec{x},\vec{p})\de^Dx\,\de^Dp$ of a function
$F:{\mathbb{R}}^{2D}\to {\mathbb{C}}$ is known to be cyclic with
respect to the star product (\ref{tre}).

A function $f(\vec{x},t)$ can be viewed as a function
$F(\vec{x},\vec{p})=f(\vec{x},\vec{x}\vec{p})$ on the $2D$
dimensional space. For such a function, the standard integral is
clearly divergent, but we can regularize it.
For each $\vec{x}$ we fix an orthonormal basis
$\{\vec{v}_j(\vec{x})\}$ for ${\mathbb{R}}^D$ with positive
orientation and with $\vec{v}_1=\vec{x}/|\vec{x}|$.
Then we consider
\begin{displaymath}
C(\vec{x})=\{\vec{u}\in {\mathbb{R}}^D\textrm{ such that }|\vec{u}\vec{v_j}|\leq
L|\vec{x}|^{-1}\;\forall\;j=1,\ldots,D\}
\end{displaymath}
which is an hypercube of edge $2L|\vec{x}|^{-1}$
and one of the edges parallel to $\vec{x}$.
The regularized integral in $\de^Dp$
is obtained restricting the integration to the interior
of $C(\vec{x})$ and normalizing
\begin{displaymath}
{\mathcal{I}}\,'(\hat{f})=\lim_{L\to
+\infty}\frac{1}{(2L)^{D-1}}\int\de^Dx\int_{C(\vec{x})}\de^D
p\,f(\vec{x},\vec{x}\vec{p})
\end{displaymath}
The normalization factor has been chosen in such a way to remove
divergencies.

In order to put ${\mathcal{I}}\,'(\hat{f})$ in explicit form
let us start by examining
the integration in $\de^Dp$. We introduce $\Lambda\in SO(D)$ as the
change of basis from the canonical one to the $\vec{v}_j$ basis.
If we make the substitution $\vec{p}\,'=\Lambda\vec{p}$, we then
find that in
the new basis $\vec{x}\vec{p}=|\vec{x}|p'_1$ and the $p$-integration
takes the form
\begin{displaymath}
\int_{C(x)} f(\vec{x},\vec{x}\vec{p})\,\de^D\!p=\int_{|p'_j|\leq L|\vec{x}|^{
-1}\;\forall\;j}f(\vec{x},|\vec{x}|p'_1)\,\de^D\!p'=\frac{(2L)^{D-1}}{|\vec{x}|^D}
\int_{-L}^{L}f(\vec{x},t)\de t
\end{displaymath}
which implies
\begin{displaymath}
{\mathcal{I}}\,'(\hat{f})=\lim_{L\to
+\infty}\int\de^Dx\frac{1}{|\vec{x}|^D}\int_{-L}^{L}
f(\vec{x},t)\de t=\int\frac{1}{|\vec{x}|^D}\,f(\vec{x},t)\,\de^D
x\,\de t
~.
\end{displaymath}
So, once again, we encounter the same choice of measure, (\ref{omog}),
which we had independently obtained in Section~3, using an analysis that
relied only on properties of \kM .

%%% ======================================================================

\section{Closing remarks}

We have here considered alternative strategies for the introduction
of an action functional for \kM .
If one is guided by the intuition that \kP\ invariance should be
a key property of the action functional then the $\mu =1$ choice
of measure is natural and cyclicity is lost.
If one is guided by the intuition that
the action functional should necessarily be
cyclic, then \kP\ invariance cannot be achieved
and one is also confronted with an unexpected
small-$x$ singularity of the measure.

Both the lack of \kP\ invariance and the small-$x$ singularity
of the measure are not necessarily alarming.
These features may not be transferred to the equations
of motion (where the physical properties of a theory should
be investigated) if one finds a cleaver way to introduce fields
in the action functional. Think for example of a ``Lagrangian
density'' of the type ${\cal L}= x \Phi(x) x \Phi(x) x \Phi(x)$,
in which case clearly  the small-$x$ singularity of the measure
would not affect the equations of motion (but clearly this
specific example
of Lagrangian density is affected by several other pathologies).
An intriguing challenge for future studies
could be the search
of \kP\ covariant non-singular equations of motion
derived from an action functional with our cyclicity-inducing
measure.

On the other hand it is tempting to attach a deeper
meaning to the peculiar small-$x$ singularity of the cyclicity-inducing
measure. In fact, already in the study of the better understood
canonical noncommutative spacetimes
we have become familiar with an unexpected source of
singularities (infrared singularities which arise through
the so-called IR/UV mixing~\cite{dougnekr,susskind,gianlucaken}).
It is therefore plausible that
also in \kM\ the small-$x$ singularity we discussed
might encode some fundamental aspect of the formalism.

\section*{Acknowledgments}
We gratefully acknowledges conversations
with G.~Mandanici.
The work of M.~A.~was supported
by a Fellowship from The Graduate School of The
University of
North Carolina. M.~A.~also thanks the Department of Physics
of the University of Rome for hospitality.

\end{document}